\documentclass[twocolumn,showpacs,showkeys]{revtex4}
\usepackage{graphicx}
\usepackage{bm}
\usepackage{color}
\usepackage{amsmath}
\usepackage{natbib}

\begin{document}

\title{Exchange effects in  magnetized quantum plasmas}
\author{Mariya Iv. Trukhanova}
\email{mar-tiv@yandex.ru}
\author{Pavel A. Andreev}
\email{andreevpa@physics.msu.ru}

\affiliation{Faculty of physics, Lomonosov Moscow State University, Moscow, Russian Federation.}

\date{\today}

\begin{abstract}
We apply the many-particle quantum hydrodynamics including the Coulomb exchange interaction to magnetized quantum plasmas. We consider a number of wave phenomenon under influence of the Coulomb exchange interaction. Since the Coulomb exchange interaction affects longitudinal and transverse-longitudinal waves we focus our attention to the Langmuir waves, Trivelpiece-Gould waves,  ion-acoustic waves in non-isothermal magnetized plasmas, the dispersion of  the longitudinal low-frequency ion-acoustic waves and    low-frequencies electromagnetic waves at $T_{e}\gg T_{i}$ . We obtained the numerical simulation  of the dispersion properties of different types of waves.
\end{abstract}

\pacs{52.30.Ex, 52.35.Dm}% PACS, the Physics and Astronomy
                             % Classification Scheme.
\keywords{quantum plasmas, exchange interaction, ion-acoustic waves, quantum hydrodynamics}
%Use showkeys class option if keyword

\maketitle

%52.30.Ex	Two-fluid and multi-fluid plasmas
%52.35.Dm	Sound waves

%52.27.Ep	Electron-positron plasmas

%%%%%%%%%%TEXT

\section{\label{sec:level1} Introduction}

The many-particle quantum hydrodynamics (MPQHD) method had been developed for the
systems of charged or neutral  particles in \cite{MaksimovTMP 1999}. General form of the quantum exchange correlations was derived there. The MPQHD for three dimensional spin 1/2 quantum plasmas with the Coulomb exchange and the spin-spin exchange interactions was obtained in 2001 in Ref. \cite{MaksimovTMP 2001 b}. Theory of ultracold quantum gases of neutral atoms, including derivation and generalization of the Gross-Pitaevskii equation, was constructed in terms of the MPQHD in 2008 \cite{Andreev PRA08}. Further development of the MPQHD with exchange interaction for low dimensional and three dimensional Coulomb quantum plasmas has been recently performed in Ref. \cite{Andreev Exchange 14}.

A quantum mechanics description for systems of N interacting  particles is based upon
the many-particle Schrodinger equation (MPSE) that specifies a wave function in a 3N-dimensional
configuration space. As wave processes, processes of information transfer, and other spin transport
processes occur in 3D physical space, it becomes necessary to turn to a mathematical method of
physically observable values that are determined in a 3D physical space.  The fundamental equations of the microscopic quantum hydrodynamics of fermions in an external electromagnetic field  had been  derived using the many-particle Schrodinger equation \cite{MaksimovTMP 2001 b}, \cite{Maksimov Izv 2000}, \cite{MaksimovTMP 2001}.  Recently there has been
an increased interest in the properties of quantum plasmas \cite{Haas PRE 00} - \cite{Brodin NJP 07}.   Dispersion relations for linear waves in amedium formed by electrons and ions and traversed by a beam of neutrons whose velocity has a nonzero constant component had been derived by methods of quantum hydrodynamics in \cite{Andreev AtPhys 08}.   The dispersion of waves, existed in the plasma in consequence of dynamic of the magnetic moments  had been investigated in \cite{Andreev IJMP 12}, \cite{pavelproc}. The instabilities at propagation of the neutron beam through the plasma  had been showed.

The  extended vorticity evolution equation for the quantum spinning plasma had been derived and  the effects of new spin forces and spin-spin interaction contributions on the motion of fermions, evolution of the magnetic moment density and vorticity generation had been predicted \cite{Trukhanova 0}.   The spin-orbital corrections to the  propagation of the whistler waves in a
astrophysical quantum magnetoplasma composed by mobile ions and electrons had been predicted in \cite{Trukhanova EPJD 13}.  The hydrodynamic model including the spin degree of freedom and the electromagnetic field had been discussed in \cite{Koide PRC 13}.  The quantum hydrodynamics  for the research of many-particles systems had been developed in \cite{Andreev PRA08}, \cite{Andreev RPJ 07}, \cite{Andreev PRB 11}.

The Coulomb exchange interactions are of great importance in many systems as well as for magnetic phenomena, and have no classical analogy.  The Coulomb exchange effects had been included in
a QHD picture  \cite{Andreev Exchange 14} for the Coulomb quantum plasmas.  To do this, the fundamental equations that determine the dynamics of functions of three variables, starting from
MPSE had been derived. This problem has been solved with the creation of a many-particle quantum hydrodynamics
 method. The contribution of the exchange interaction in the dispersion of the Langmuir \cite{Andreev Exchange 14} and ion-acoustic waves for three and  two dimensional quantum plasmas  had been shown. It had been derived that the exchange interaction between particles with same spin direction and particles with opposite spin directions are different.    Recently, the  kinetic plasma model containing fermion exchange effects were investigated in \cite{Zamanian PRE 13 exch} and the influence of exchange effect on low frequency dynamics, in particular ion acoustic waves was predicted.  The generalization of the Vlasov equation to include exchange effects was presented allowing for electromagnetic mean fields and the correction to classical Langmuir waves in plasmas was found in \cite{Zamanian 14 exch}.

The influence of electron-exchange and quantum screening on the collisional entanglement fidelity for the elastic electron–ion collision was investigated  \cite{Hong PScr 14}. The effective Shukla$ -$ Eliasson potential and the partial wave method had been used to obtain the collisional entanglement fidelity in quantum plasmas as a function of the electron-exchange parameter, Fermi energy, plasmon energy and collision energy.   The electron-exchange effects on the charge capture process had been investigated in degenerate quantum plasmas \cite{Jung PP 14}. It had been showed that the electron-exchange effect enhances the charge capture radius and the charge capture cross section in semiconductor quantum plasmas.

This paper is organized as follows.  In Sec.  (\ref{sec:level1})  we
present the quantum hydrodynamics equations taken into account the Coulomb exchange interactions in a selfconsistent
field approximation.    In Sec. (\ref{sec:level2}) we presented the dispersion characteristics of high frequency  Langmuir
waves and low frequency waves  with the account of Coulomb exchange interactions. In Sect. (\ref{sec:level3}) we show the influence of exchange interactions on the dispersion of ion-acoustic waves in non-isothermal magnetized plasmas.   In Sec. (\ref{sec:level4})  we  show
that exchange interactions  leads to the existence of
a dispersion characteristics of   the Trivelpiece-Gould waves. In Sec.  (\ref{sec:level5}) and Sec.  (\ref{sec:level6})   we show the influence of exchange interactions on the dispersion of  the longitudinal low-frequency ion-acoustic waves and    low-frequencies electromagnetic waves at $T_{e}\gg T_{i}$ in the magnetized plasma.

\section{\label{sec:level1} Model}

  We apply the equations for the system
of charged particles in the external magnetic field  \cite{Andreev Exchange 14}. For a 3D system of particles the continuity equations and the momentum
balance equations for electrons and ions may be written down in terms of
electrical intensity of the field that is created by charges
 of the particle system. Thus  the continuity equations
for ions and electrons are

\begin{equation}\label{EXCH_SEC cont eq electrons}
\partial_{t}n_{e}+\nabla(n_{e}\textbf{v}_{e})=0, \end{equation}
\textit{and}
\begin{equation}\label{EXCH_SEC cont eq ions}
\partial_{t}n_{i}+\nabla(n_{i}\textbf{v}_{i})=0. \end{equation}

The equation of motion for electrons
 is

$$m_{e}n_{e}(\partial_{t}+\textbf{v}_{e}\nabla)\textbf{v}_{e}+\nabla p_{e}-\frac{\hbar^{2}}{4m_{e}}n_{e}\nabla\Biggl(\frac{\triangle n_{e}}{n_{e}}-\frac{(\nabla n_{e})^{2}}{2n_{e}^{2}}\Biggr)$$
\begin{equation}\label{EXCH_SEC Euler eq electrons non INT} =q_{e}n_{e}\biggl(\textbf{E}_{ext}+\textbf{E}_{int}+\frac{1}{c}[\textbf{v}_{e},\textbf{B}_{ext}]\biggr)+2^{4/3} q_{e}^{2}\sqrt[3]{\frac{3}{\pi}}\sqrt[3]{n_{e}}\nabla n_{e},\end{equation}

\textit{and} the Euler equation for ions

$$m_{i}n_{i}(\partial_{t}+\textbf{v}_{i}\nabla)\textbf{v}_{i}+\nabla p_{i}-\frac{\hbar^{2}}{4m_{i}}n_{i}\nabla\Biggl(\frac{\triangle n_{i}}{n_{i}}-\frac{(\nabla n_{i})^{2}}{2n_{i}^{2}}\Biggr)$$
\begin{equation}\label{EXCH_SEC Euler eq ions non INT} =q_{i}n_{i}\biggl(\textbf{E}_{ext}+\textbf{E}_{int}+\frac{1}{c}[\textbf{v}_{i},\textbf{B}_{ext}]\biggr)+2^{4/3} q_{i}^{2}\sqrt[3]{\frac{3}{\pi}}\sqrt[3]{n_{i}}\nabla n_{i}.\end{equation}

The second
terms on the left sides of Eq. (\ref{EXCH_SEC Euler eq electrons non INT}) and (\ref{EXCH_SEC Euler eq ions non INT}) are the gradient of the thermal pressure or the
Fermi pressure for degenerate electrons and ions. It appears
as the thermal part of the momentum flux related
to distribution of particles on states with different momentum.
The third terms are the quantum Bohm potential
appearing as the quantum part of the momentum
flux. In the right-hand sides of the Euler equations we
present interparticle interaction and interaction of particles
with external electromagnetic fields. The first group
of terms in the right-hand side of the Euler equations
describe interaction with the external electromagnetic
fields.
The force fields of Coulomb exchange interaction of electrons and positrons (the second terms on the right side of (\ref{EXCH_SEC Euler eq electrons non INT}) and (\ref{EXCH_SEC Euler eq ions non INT})) are obtained for fully polarized systems of identical particles. For polarized systems of electrons or ions equations
of state appears $p_{a,3D\uparrow\uparrow}=2^{2/3}\hbar^2n^{5/3}_{a,3D}/5m_a$ for 3D mediums, where subindex $\uparrow\uparrow$ means that all particles
have same spin direction. The ratio of
polarizability $\eta=\frac{\mid
n_{\uparrow}-n_{\downarrow}\mid}{n_{\uparrow}+n_{\downarrow}}$,
with indexes $\uparrow$ and $\downarrow$ means particles with spin
up and spin down. In general case of partially polarized system of particles
we can write $p_{a,3D\Updownarrow}=\vartheta_{3D}(3\pi^{2})^{2/3}\hbar^{2}n_{a,3D}^{5/3}/(5m_{a})$
for partially polarized systems, that
means that part of states contain two particle with opposite
spins and other occupied states contain one particle
with same spin direction    \cite{Andreev Exchange 14}
\begin{equation}\label{EXCHANGE} \vartheta_{3D}=\frac{1}{2}[(1+\eta)^{5/3}+(1-\eta)^{5/3}],\end{equation}

For partially polarized particles the force fields reappear as $\textbf{F}_{C,a(3D)}=\zeta_{3D} q_{a}^{2}\sqrt[3]{\frac{3}{\pi}}\sqrt[3]{n_{a}}\nabla n_{a}$, with
\begin{equation}\label{EXCH_SEC} \zeta_{3D}=(1+\eta)^{4/3}-(1-\eta)^{4/3}\end{equation}
We should mention that coefficient $\zeta_{3D}\sim\eta$ is proportional to spin polarization. Limit cases of $\zeta_{3D}$ are $\zeta_{3D}(0)=0$, $\zeta_{3D}(1)=2^{4/3}$.

Considering two electrons one finds that full wave function
is anti-symmetric. If one has two electrons with parallel
spins one has that wave function is symmetric on
spin variables, so it should be anti-symmetric on space
variables. In opposite case of anti-parallel spins one has
anti-symmetry of wave function on spin variables and
symmetry of wave function on space variables.

Systems of unpolarized electrons then average numbers
of electrons with different direction of spins equal to
each other, we find that average number of particles for
a chosen with parallel and anti-parallel spins is the same.
Consequently we have that average exchange interaction
equals to zero.
In partly polarized systems the numbers of particles
with different spin are not the same. In this case a contribution
of the average exchange interaction appears.
At full measure it reveals in fully polarized system then
all electrons have same direction of spins. In accordance
with the previous discussion we find that exchange interaction,
for this configuration, gives attractive contribution
in the force field.

\section{\label{sec:level2} Applications}

In this section we consider small perturbations of equilibrium state describing by nonzero particle concentration $n_{0}$, and zero velocity field $\textbf{v}_{0}=0$ and electric field $\textbf{E}_{0}=0$.

Assuming that perturbations are monochromatic
\begin{equation}\label{EXCH_SEC perturbations}
\left(\begin{array}{ccc} \delta n_{e}
 \\
\delta n_{i}
 \\
\delta \textbf{v}_{e}
 \\
\delta \textbf{v}_{i}
 \\
\delta \textbf{E}
\\
\delta \textbf{B}
\\
\end{array}\right)=
\left(\begin{array}{ccc}
N_{Ae} \\
N_{Ai} \\
\textbf{V}_{Ae} \\
\textbf{V}_{Ai} \\
\textbf{E}_{A}\\
\textbf{B}_{A}\\
\end{array}\right)e^{-\imath\omega t+\imath \textbf{k} \textbf{r}},\end{equation}
we get a set of linear algebraic equations relatively to $N_{A}$ and $V_{A}$. Condition of existence of nonzero solutions for amplitudes of perturbations gives us a dispersion equation in the form of

\begin{figure}[bh]
\includegraphics[width=9cm,angle=0]{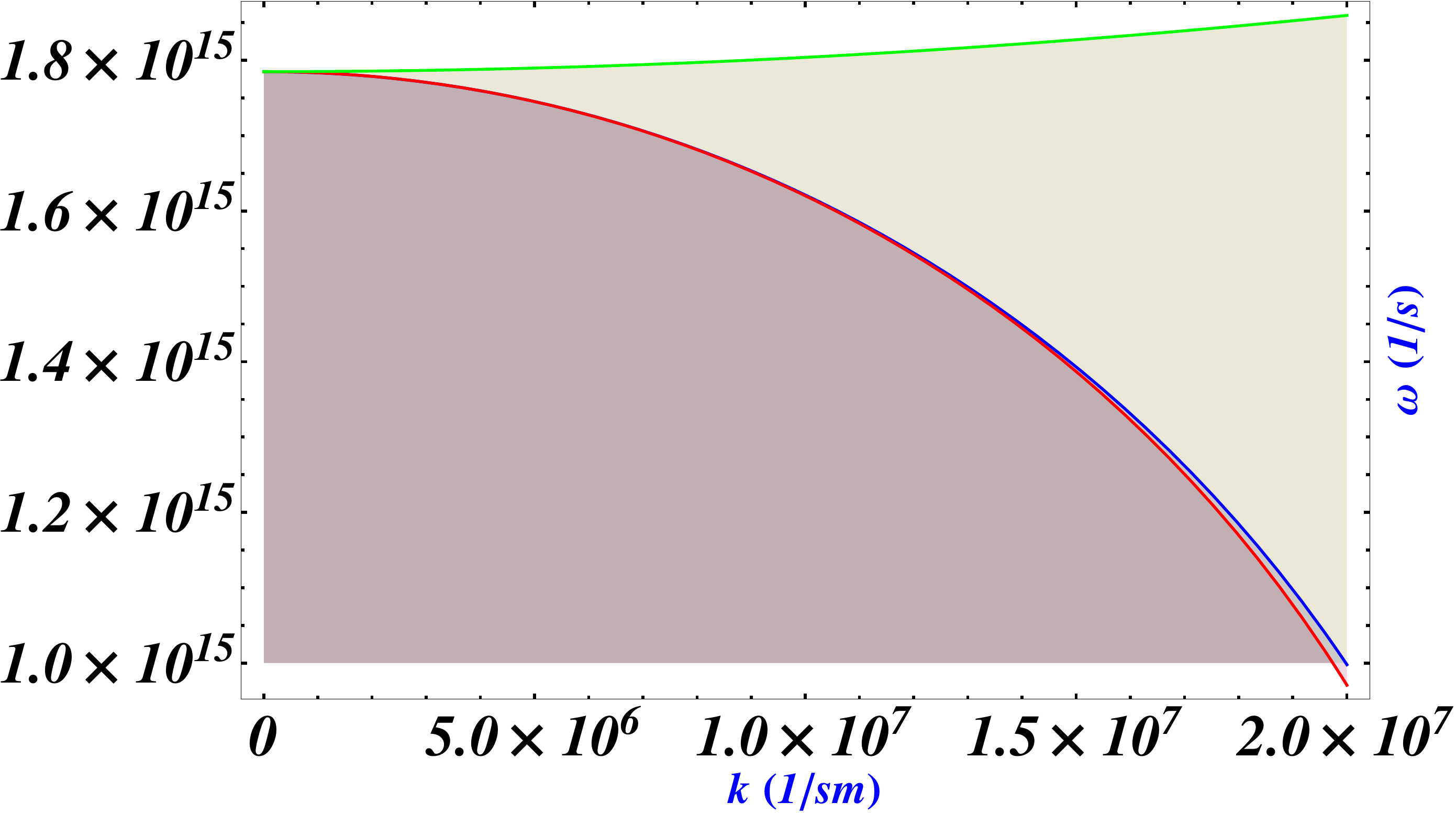}
\caption{\label{exw22}\em The figure shows the dispersion characteristic of the
quantum Langmuir wave frequency $\omega$ versus the wave
vector $k$, which is described by the equation (\ref{EXCH_SEC disp Lang 3D}). The {\color{green}green branch} of dispersion is presented the classical high frequency Langmuir wave, the {\color{red}red} and {\color{blue}blue branches} characterize the Coulomb exchange interactions and quantum Bohm potential influence, where $n_{0e}\simeq10^{21}cm^{-3}$, $\eta=1$.}   \label{Langmir}  \end{figure}
 \begin{figure}[bh]
\includegraphics[width=9cm,angle=0]{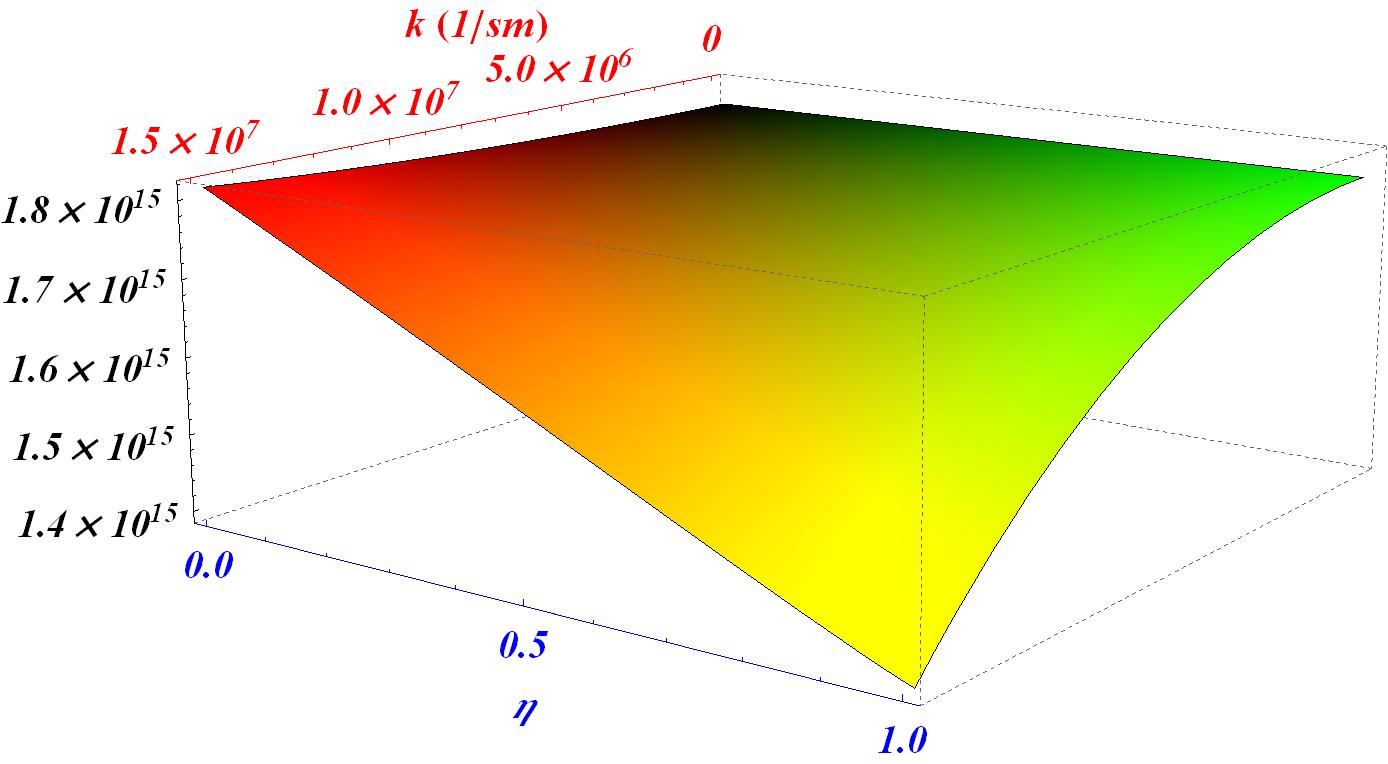}
\caption{\label{exw22}\em The figure shows the dispersion characteristic of the
quantum Langmuir wave frequency $\omega$ as a function of  the wave
vector $k$ and the ratio of
polarizability $\eta$, which is described by the equation (\ref{EXCH_SEC disp Lang 3D}).  The Coulomb exchange interactions and quantum Bohm potential are taken into account and  $n_{0e}\simeq10^{21}cm^{-3}$.}   \label{Langmir2}
     \end{figure}
\begin{equation}\label{EXCH_SEC Langmuir freq 3D} \omega_{Le,3D}^{2}=\frac{4\pi e^2 n_{0,3D}}{m_{e}}.\end{equation}

$$\omega^{2}=\omega_{Le, 3D}^{2}-\zeta_{3D}\sqrt[3]{\frac{3}{\pi}}\frac{e^{2}}{m_{e}}\sqrt[3]{n_{0e}}k^{2}$$
\begin{equation}\label{EXCH_SEC disp Lang 3D} +\vartheta_{3D}\frac{(3\pi^{2})^{2/3}\hbar^{2}n_{0e}^{2/3}}{3m_{e}^{2}}k^{2}+\frac{\hbar^{2}k^{4}}{4m_{e}^{2}}, \end{equation}
  where the dispersion of quantum Langmuir
waves with the account of Coulomb exchange interactions (\ref{EXCH_SEC disp Lang 3D}) is presented at Fig. (\ref{Langmir}). In general case of of partially polarized system of particles with spin up and spin down, the dispersion of quantum Langmuir
waves as a function of ratio of
polarizability with account of Coulomb exchange interactions (\ref{EXCH_SEC disp Lang 3D}) is presented at Fig. (\ref{Langmir2}).

We see that the first term in Eq. (\ref{EXCH_SEC disp Lang 3D}) is proportional to the electron equilibrium concentration $\sim n_{0e}$ and grows faster then the second term   $\sim n_{0e}^{1/3}$. The third term has an intermediate rate of grow being proportional to $\sim n_{0e}^{2/3}.$  The Coulomb exchange interaction is larger than the Fermi pressure when $\sim n_{0e}\leq10^{24}cm^{-3},$  this situation is realized in metals and semiconductors, but in the astrophysical objects like white drafts   $\sim n_{0e}\leq10^{28}cm^{-3},$  the Fermi pressure is larger than the Coulomb exchange interaction.

We have considered high frequency waves. Nest step is consideration of the low frequency excitations
$$\omega_{3D}(k)=kv_{s,3D}\sqrt{1-\frac{\zeta_{3D}}{\vartheta_{3D}}\frac{3}{4\pi}\sqrt[3]{\frac{3}{\pi}}\frac{\omega_{Le,3D}^{2}}{n_{0e,3D}^{2/3}v_{Fe,3D}^{2}}}\times$$
\begin{equation}\label{EXCH_SEC disp ion-acoustic 3D}   \times\frac{1}{\sqrt{1+(kr_{De,3D})^{2} \biggl(1-\frac{\zeta_{3D}}{\vartheta_{3D}}\frac{3}{4\pi}\sqrt[3]{\frac{3}{\pi}}\frac{\omega_{Le,3D}^{2}}{n_{0e,3D}^{2/3}v_{Fe,3D}^{2}}\biggr)}},\end{equation}
where $v_{s,3D}=\sqrt{m_{e}/m_{i}}\sqrt{\vartheta_{3D}}\cdot v_{Fe,3D}/3$ is the three dimensional velocity of sound, $r_{De,3D}=\sqrt{\vartheta_{3D}}v_{Fe,3D}/(3\omega_{Le,3D})$ is the Debye radius. In formula (\ref{EXCH_SEC disp Lang 3D}) and similar formulas below we extract contribution of the Fermi pressure. Hence formulas for ion-acoustic waves contains well-known contribution of the pressure multiplied by factor showing contribution of exchange interaction.

In the long wavelength limit we have
\begin{equation}\label{EXCH_SEC IAW 3D LWL} \omega(k)=kv_{s,3D}\sqrt{1-\frac{\zeta_{3D}}{\vartheta_{3D}}\frac{3}{4\pi}\sqrt[3]{\frac{3}{\pi}}\frac{\omega_{Le,3D}^{2}}{n_{0e,3D}^{2/3}v_{Fe,3D}^{2}}}.\end{equation}

In the short wavelength limit we find
\begin{equation}\label{EXCH_SEC IAW 3D SWL} \omega^{2}(k)=\omega_{Li,3D}^{2}. \end{equation}

\section{\label{sec:level3} Ion-acoustic waves in non-isothermal magnetized plasmas}

            \begin{figure}[h]
\includegraphics[width=9cm,angle=0]{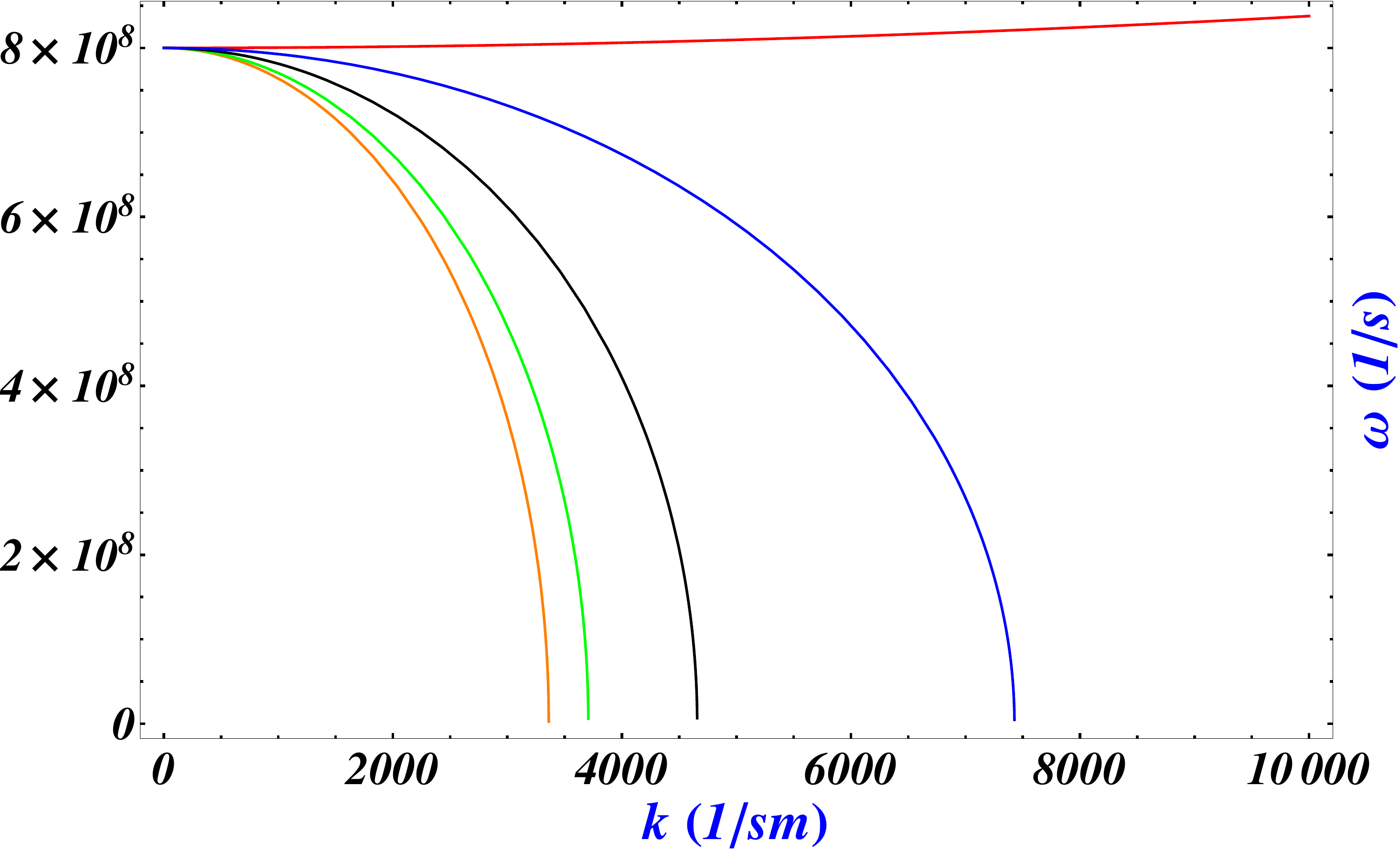}
\caption{\label{sw21}\em The figure shows the dispersion characteristic of the long wavelength
ion-acoustic wave   frequency $\omega$ versus the wave vector $k$, which
is described by the equation (\ref{EXCH_SEC1}). The {\color{red}red mode} shows the dispersion characteristic of the long wavelength ion-acoustic wave, where the thermal speed is defined by the Fermi pressure, the {\color{yellow}orange mode} presents the  Coulomb exchange interactions influence for $\eta=1$, the {\color{green}green mode} for the case of partially polarized system $\eta=0.8$, the {\color{black}black mode} for $\eta=0.5$ and   the {\color{blue}blue mode} presents the  Coulomb exchange interactions influence for $\eta=0.2$.    System
parameters are assumed to be as follows: $B_0\simeq5\cdot10^{5}$ G - the uniform magnetic field, $n_0\simeq10^{18}$cm$^{-3}$ - the equilibrium density.}
     \end{figure}      \begin{figure}[h]
\includegraphics[width=9.6cm,angle=0]{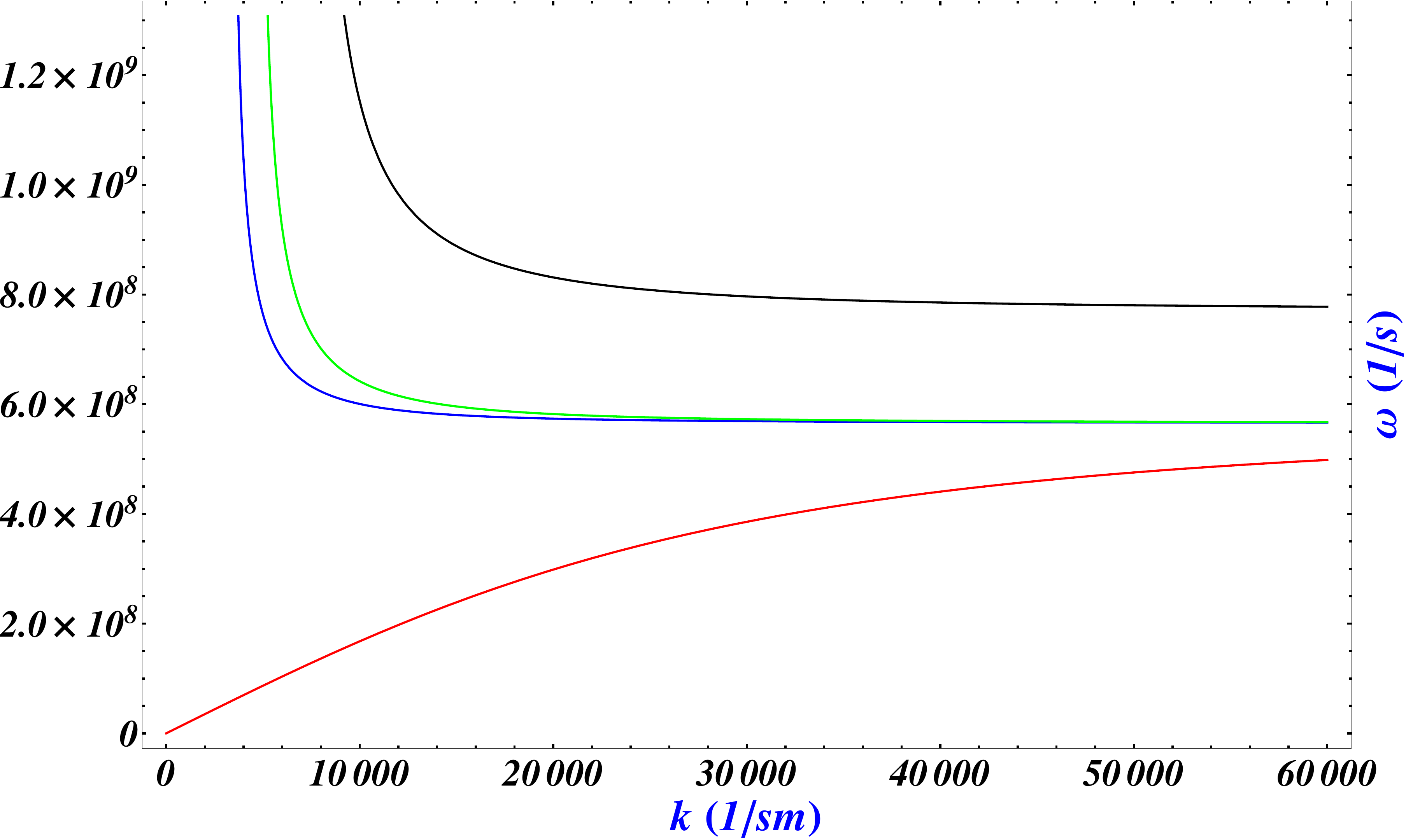}
\caption{\label{sw22}\em The figure shows the dispersion characteristic of the long wavelength
ion-acoustic wave   frequency $\omega$ versus the wave vector $k$, which
is described by the equation (\ref{EXCH_SEC2}). The {\color{red}red mode} shows the dispersion characteristic of the long wavelength ion-acoustic wave, where the thermal speed is defined by the Fermi pressure and the {\color{blue}blue mode} presents the  Coulomb exchange interactions influence for $\eta=1$, the {\color{green}green mode} for $\eta=0.5$ and the black mode for  $\eta=0.2$.    System
parameters are assumed to be as follows: $B_0\simeq5\cdot10^{5}$ G - the uniform magnetic field, $n_0\simeq10^{18}$cm$^{-3}$ - the equilibrium density.}
     \end{figure}

   \begin{figure}[h]
\includegraphics[width=9.4cm,angle=0]{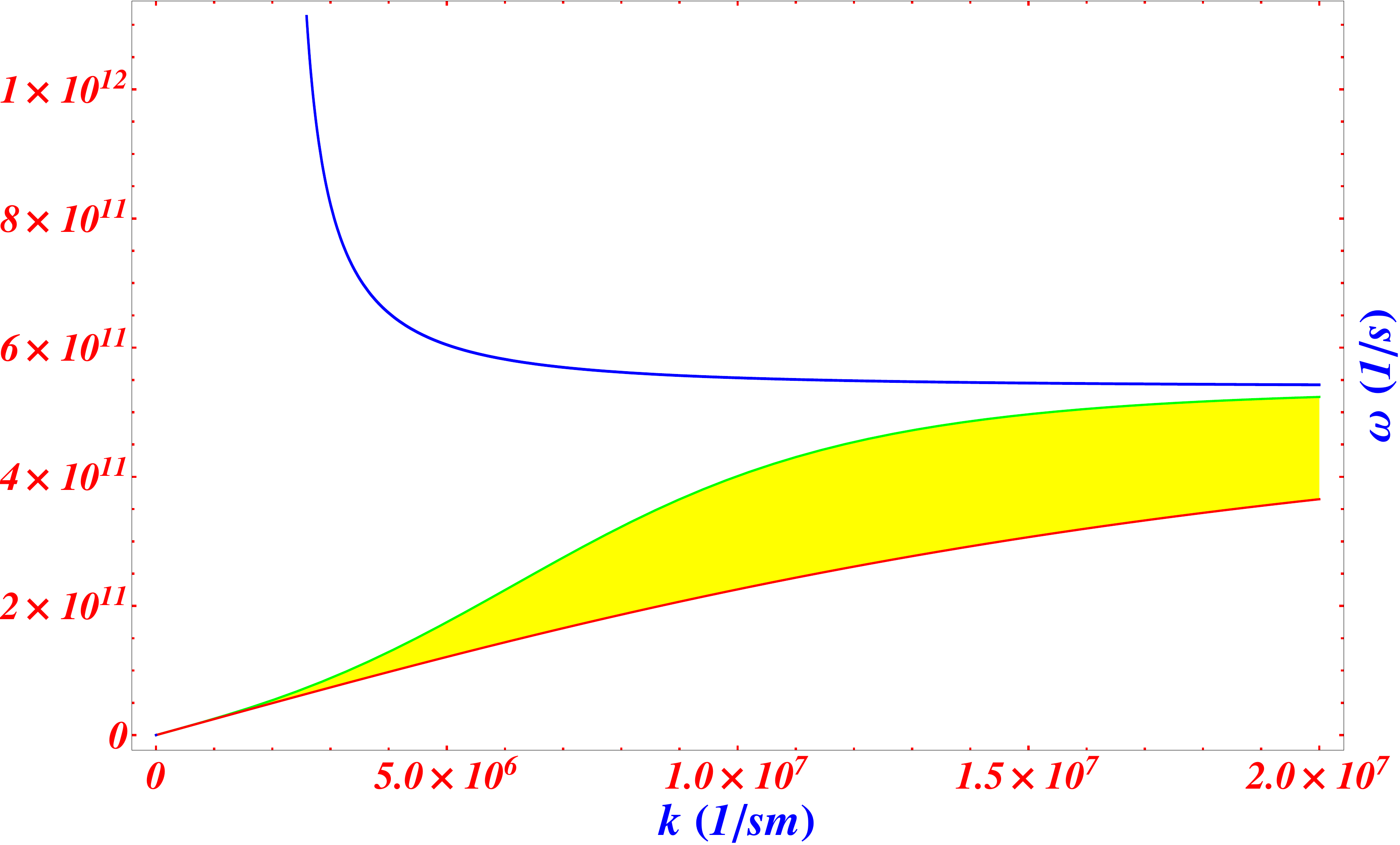}
\caption{\label{ssw21}\em The figure shows the dispersion characteristic of the short wavelength
quantum Langmuir wave frequency $\omega$ versus the wave
vector k, which is described by the equation (\ref{EXCH_SEC disp Lang 3D}). The {\color{green}green mode}  shows the
dispersion characteristic of quantum ion-acoustic waves which
are described by equation (\ref{EXCH_SEC1}), the {\color{red}red mode} shows the
dispersion characteristic of classic ion-acoustic waves which occur
from (\ref{EXCH_SEC1}) if the term proportional to ~$\hbar^2$ is neglected. The {\color{blue}blue mode} shows the
dispersion characteristic of the short wavelength wave  where the Coulomb exchange interactions are dominate.}
     \end{figure}

            Lets discuss the ion-acoustic waves in non-isothermal magnetized plasmas with  $T_{e}>>T_{i}\approx0$ at $\Omega_{e}^{2}\gg\omega_{Li}^{2}\gg\Omega_{i}^{2}$ \cite{Kuzelev Ruhadze UFN1999}. In these conditions the ion-acoustic waves exist giving two branches of wave dispersion. One branch is at $\omega<\Omega_{i}$ and another one at $\omega>\Omega_{i}$.

Corresponding dispersion equation appears as
\begin{equation}\label{EXCH_SEC0} 1-\frac{k_{\perp}^{2}\omega_{Li}^{2}}{k^{2}(\omega^{2}-\Omega_{i}^{2})}-\frac{k_{z}^{2}\omega_{Li}^{2}}{k^{2}\omega^{2}}+\frac{\omega_{Le}^{2}}{k^{2}v_{Te}^{2}}=0, \end{equation}
where $v_{Te}^{2}$  - is the thermal electron velocity.
Using the approximation $\omega_{Li}^{2}\gg\Omega_{i}^{2}$ and $\theta\neq0$ we find
\begin{equation}\label{EXCH_SEC1} \omega_{+}^{2}=\frac{\omega_{Li}^{2}+\Omega_{i}^{2}(1+\omega_{Li}^{2}/k^{2}v_{s}^{2})}{1+\omega_{Li}^{2}/k^{2}v_{s}^{2}}, \end{equation}
and
\begin{equation}\label{EXCH_SEC2} \omega_{-}^{2}=\frac{\omega_{Li}^{2}\Omega_{i}^{2}\cos^{2}\theta}{\omega_{Li}^{2}+\Omega_{i}^{2}(1+\omega_{Li}^{2}/k^{2}v_{s}^{2})},\end{equation}

where

  $$\upsilon^2_s=-\xi_{3D}\frac{e^2}{m_i}\sqrt[3]{\frac{3}{\pi}}\sqrt[3]{n_{0e}}$$       \begin{equation}\label{v} \qquad\qquad\qquad+\vartheta_{3D}\frac{(3\pi^2)^{2/3}\hbar^2n_{oe}^{2/3}}{3m_em_i}+\frac{\hbar^2k^2}{4m_em_i}.\end{equation}
contains the Coulomb exchange interaction between electrons (the first term).
The second term appears from the Fermi pressure.
and  the third  term describes contribution of the quantum
Bohm potential

Solutions of the equation (\ref{EXCH_SEC0}) for the different ratio of polarizability are presented by two
branches of the dispersion  at Fig. (\ref{sw21}), (\ref{sw22}) and (\ref{ssw21}).

When the distribution of the wave is parallel to the background
magnetic field $\theta=0$      the ion-acoustic   modes have the form of
\begin{equation}  \label{w23}   \omega^2=\Omega^2_{i}\end{equation}
                                 and
                  \begin{equation}  \label{w23}   \omega^2=\Biggl(-\xi_{3D}\frac{e^2}{m_i}\sqrt[3]{\frac{3}{\pi}}\sqrt[3]{n_{0e}}+\vartheta_{3D}\frac{(3\pi^2)^{2/3}\hbar^2n_{oe}^{2/3}}{3m_em_i}+\frac{\hbar^2k^2}{4m_em_i}\Biggr)\times\end{equation}

             \[   \times\frac{\omega^2_{Li}k^2}{\omega^2_{Li}+k^2(-\xi_{3D}\frac{e^2}{m_i}\sqrt[3]{\frac{3}{\pi}}\sqrt[3]{n_{0e}}+\vartheta_{3D}\frac{(3\pi^2)^{2/3}\hbar^2n_{oe}^{2/3}}{3m_em_i}+\frac{\hbar^2k^2}{4m_em_i})}
               \]

\section{\label{sec:level4} The Trivelpiece-Gould waves}

Trivelpiece-Gould wave \cite{Kuzelev Ruhadze UFN1999}, \cite{Tercas PP 08} is a longitudinal wave appearing along with the Langmuir wave at propagation of waves at an angle to external magnetic field $k_{x}\neq0$ and $k_{z}\neq0$. It is a low-frequency wave with $\omega\ll\mid\Omega_{e}\mid$. It exists at long wavelengths limit $\ll1$, $k_{z}^{2}v_{Te}^{2}\ll\omega^{2}$. All of these conditions give the following dispersion equation
\begin{equation}\label{EXCH_SEC6} 1-\frac{\omega_{Le}^{2}}{\omega^{2}}\frac{k_{z}^{2}}{k^{2}}+\frac{\omega_{Le}^{2}}{k^{2}v_{Te}^{2}}\frac{k_{x}^{2}v_{Te}^{2}}{\Omega_{e}^{2}}=0\end{equation}

Dispersion dependence of the Trivelpiece-Gould wave in the classical magnetic  field  appears as  \cite{Tercas PP 08}

\begin{equation}\label{EXCH_SEC7} \omega^{2}=\frac{\omega_{Le}^{2}\cos^{2}\theta}{1+\frac{\omega_{Le}^{2}\sin^{2}\theta}{\Omega_{e}^{2}}}.\end{equation}

It looks like thermal velocity does not affect this spectrum. If it is correct we can mention that the exchange interaction does not influence it either.
In the quantum  external magnetic field the thermal velocity effect might be important. This
corresponds to a regime of very strong magnetic field in which the external field strength
approaches or exceeds the quantum critical magnetic field, $B_0\geq4.4\cdot10^{13}$ G. Evidently, since $\hbar\Omega_{e}>>T_e$ there must exist the
frequency

\begin{equation}\label{EXCH_SEC8} \omega^{2}=\frac{m_e\Omega_{e}\tan^{2}\theta}{\hbar}\times
\end{equation}
\[
\times\Biggl(-\xi_{3D}\frac{e^2}{m_e}\sqrt[3]{\frac{3}{\pi}}\sqrt[3]{n_{0e}}+\vartheta_{3D}\frac{(3\pi^2)^{2/3}\hbar^2n_{oe}^{2/3}}{3m_e^2}+\frac{\hbar^2k^2}{4m_e^2}\Biggr).
\]
%\begin{equation}\label{EXCH_SEC}\end{equation}

\section{\label{sec:level5} The longitudinal low-frequency ion-acoustic waves in  magnetized plasmas}

In absence of external magnetic field longitudinal oscillations exists in electron-ion plasmas with hot electrons and cold ions \cite{Kuzelev Ruhadze UFN1999}.
They are weakly damping ion-acoustic oscillations \cite{Ahiezer}. Dispersion of these waves is
\begin{equation}\label{EXCH_SEC ion-ac osc} \omega_{s}(\textbf{k})=\frac{kv_{s}}{\sqrt{1+k^{2}r_{D}^{2}}}, \end{equation}
where $r_{D}=\sqrt{T_{e}/4\pi e^{2}n_{0}}$ is the Debay radius, $v_{s}=\sqrt{T_{e}/m_{i}}$. This solution corresponds to phase velocities of waves, which are intermediate in compare with other parameters of plasmas $v_{i}\ll\omega/k\ll v_{e}$ with the thermal velocities of electrons $v_{e}=\sqrt{T_{e}/m_{e}}$ and ions $v_{i}=\sqrt{T_{i}/m_{i}}$.

If equilibrium state of a medium reveals distribution of electrons with non-zero magnetization, as it happens in ferromagnetic domains, the Coulomb exchange interaction gives considerable contribution in spectrum of the longitudinal waves \cite{Andreev Exchange 14}.

It is well-known that an external magnetic field affects the ion-acoustic waves at $\omega_{s}(\textbf{k})\leq \Omega_{i}$ and $k\rho_{i}\leq 1$.
Presence of external magnetic field creates or increases difference in occupation of spin-up and spin-down states. Consequently, account of the Coulomb exchange interaction in magnetized plasmas is even more important than in plasmas with no external magnetic field.

Under conditions  $v_{i}\ll\mid\omega/k_{\parallel}\mid\ll v_{e}$  we find the following dispersion equation  for the     longitudinal low-frequency ion-acoustic waves

\begin{equation}\label{EXCH_SEC9} 1+\frac{\omega_{Le}^{2}}{\Omega_{e}^{2}}+\frac{\omega_{Li}^{2}}{k^{2}v_{s}^{2}}
-\frac{\omega_{Li}^{2}}{\omega^{2}}\cos^{2}\theta-\frac{\omega_{Li}^{2}}{\omega^{2}-\Omega_{i}^{2}}\sin^{2}\theta=0, \end{equation}
where $\Omega_{a}=\frac{q_{a}B_{0}}{m_ac}$ - are the the electron or ion
 cyclotron  frequency.

Increasing of the Coulomb exchange interaction in compare with the Fermi pressure can break condition of the ion-acoustic wave existence $v_{i}\ll\mid\omega/k_{\parallel}\mid\ll v_{e}$.

Getting into account the fact that in the problem under consideration the second term in equation (\ref{EXCH_SEC9}) much smaller than the third term we find next solution
\begin{equation}\label{EXCH_SEC longit waves in magn field} \omega^{2}(k,\theta)=\frac{1}{2}(\omega_{s}^{2}+\Omega_{i}^{2}) \pm\frac{1}{2}\sqrt{(\omega_{s}^{2}+\Omega_{i}^{2})^{2}-4\omega_{s}^{2}\Omega_{i}^{2}\cos^{2}\theta}.\end{equation}
We have also neglected small anisotropic terms.

In presence of an external magnetic field, there are two longitudinal low-frequency oscillations instead of the ion-acoustic wave.

At $k\rightarrow0$ we find from formula (\ref{EXCH_SEC longit waves in magn field})
\begin{equation}\label{EXCH_SEC10} \omega(k,\theta)=kv_{s}\cos\theta,\end{equation}
and
\begin{equation}\label{EXCH_SEC11}\omega(k,\theta)=\Omega_{i}\biggl(1+\frac{k^{2}v_{s}^{2}\sin^{2}\theta}{2\Omega_{i}^{2}}\biggr).\end{equation}

\begin{figure} [h]
   \centering\begin{tabular}{c}
     \includegraphics [scale=0.24] {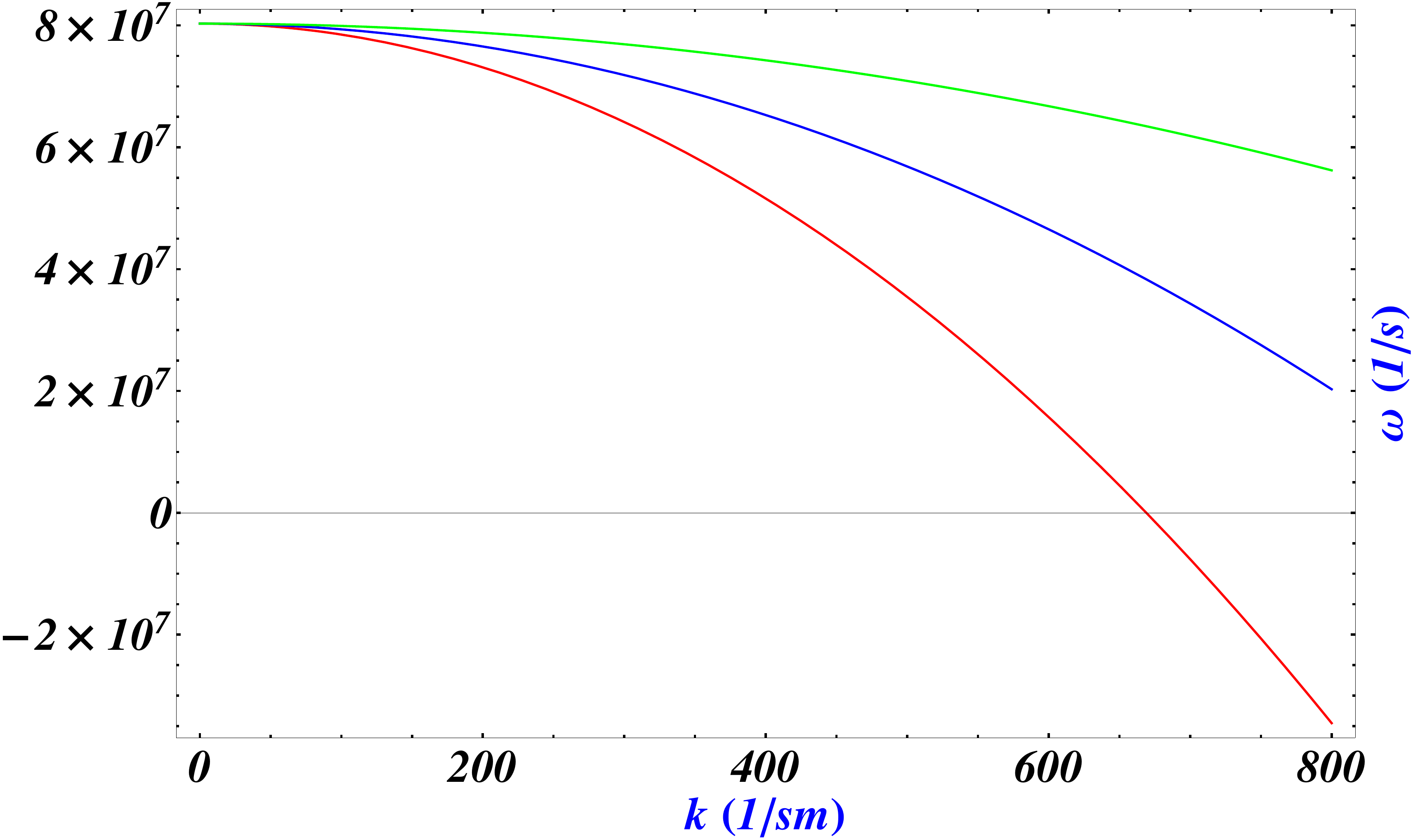} \\\end{tabular}
\caption{{\em The figure shows the dispersion characteristic of the low-frequency oscillations in the long wavelengths $k\rightarrow\infty$ limit, which is described by the equation (\ref{EXCH_SEC11}), $n_0\simeq 10^{18} sm^{-3}$, $B_0=5\cdot10^4$ G. The {\color{red}red branch} characterizes the total polarized system $\eta=1$, the {\color{blue}blue mode} $\eta=0.5$ and {\color{green}green mode} $\eta=0.2$ present the dispersion properties of  partially polarized systems. }
    }\label{ex5} \end{figure}

The condition for instability is thus that the  negative term of $v_{s}^2$ dominates over all the others. When the Coulomb exchange interactions is larger than Fermi pressure $n_0\leq 10^{23} sm^{-3}$, the solution  (\ref{EXCH_SEC10})  is instability. The solution (\ref{EXCH_SEC11}) is presented at Fig. (\ref{ex5}) as a function of $\eta$.

In opposite limit of small wavelengths $k\rightarrow\infty$, or in other terms $kr_{D}\gg1$, from formula (\ref{EXCH_SEC longit waves in magn field}) we obtain
\begin{equation}\label{EXCH_SEC longit waves in magn field Large k}\omega_{\pm}^{2}(\theta)=\frac{1}{2}(\omega_{Li}^{2}+\Omega_{i}^{2})
\pm\frac{1}{2}\sqrt{(\omega_{Li}^{2}+\Omega_{i}^{2})^{2}-4\omega_{Li}^{2}\Omega_{i}^{2}\cos^{2}\theta}.\end{equation}
Formula (\ref{EXCH_SEC longit waves in magn field Large k}) is obtained at $k^{2}\rho_{i}^{2}\ll1$. It can be applied at $\frac{T_{i}}{T_{e}}\frac{\omega_{Li}^{2}}{\Omega_{i}^{2}}\ll1$. Under condition $\omega_{Li}\gg\Omega_{i}$ formula (\ref{EXCH_SEC longit waves in magn field Large k}) does not work. Formula (\ref{EXCH_SEC longit waves in magn field}) can be applied in the long wavelength limit $kr_{D}\ll1$ if condition $\omega_{Li}\gg\Omega_{i}$ is satisfied.

Under conditions $kv_{s}\gg\Omega_{i}$ and $kr_{D}\ll1$, we obtain the following solutions from formula (\ref{EXCH_SEC longit waves in magn field})
\begin{equation}\label{EXCH_SEC longit waves in magn field Small k Dense up} \omega(k)=kv_{s},\end{equation}
and
\begin{equation}\label{EXCH_SEC longit waves in magn field Small k Dense dawn} \omega(k,\theta)=\Omega_{i}\cos\theta.\end{equation}

The Coulomb exchange interaction in (\ref{EXCH_SEC longit waves in magn field Small k Dense up}) is larger than the Fermi pressure when $\sim n_{0e}\leq10^{25}cm^{-3},$  this situation is realized in metals and semiconductors, but in the astrophysical objects like white drafts   $\sim n_{0e}\leq10^{28}cm^{-3},$  the Fermi pressure is larger than the Coulomb exchange interaction.

At $k\rightarrow0$ ($kv_{s}\ll\Omega_{i}$) larger of solutions (\ref{EXCH_SEC longit waves in magn field}) presented by formula (\ref{EXCH_SEC longit waves in magn field Small k Dense up}) getting to $\Omega_{i}$.

We can also present corresponding refractive index
\begin{equation}\label{EXCH_SEC12} N^{2}=\frac{c^{2}}{\upsilon_{s}^{2}}\frac{\omega^{2}-\Omega_{i}^{2}}{\omega^{2}-\Omega_{i}^{2}\cos^{2}\theta}.\end{equation}
%\begin{figure} [htbp]
 %  \centering\begin{tabular}{c}
  %   \includegraphics [scale=0.3] {N.pdf} \\\end{tabular}
%\caption{{\em The figure shows the corresponding classical refractive index (\ref{EXCH_SEC12}) in the limit of the long wavelength limit $kr_{D}\ll1$, $n_0\simeq 10^{18} sm^{-3}$, $B_0=5\cdot10^4$ G. The blue branch is the index $c^2/\upsilon_{s}^{2}$, and the thermal speed is determined by the Fermi pressure  $\upsilon_{s}^{2}\sim\upsilon_{Fe}^{2}$.}
 %   }\label{N} \end{figure}

%\centering\begin{tabular}{c}\end{tabular}
   \begin{figure} [h]
   \centering\begin{tabular}{c}
     \includegraphics [scale=0.29] {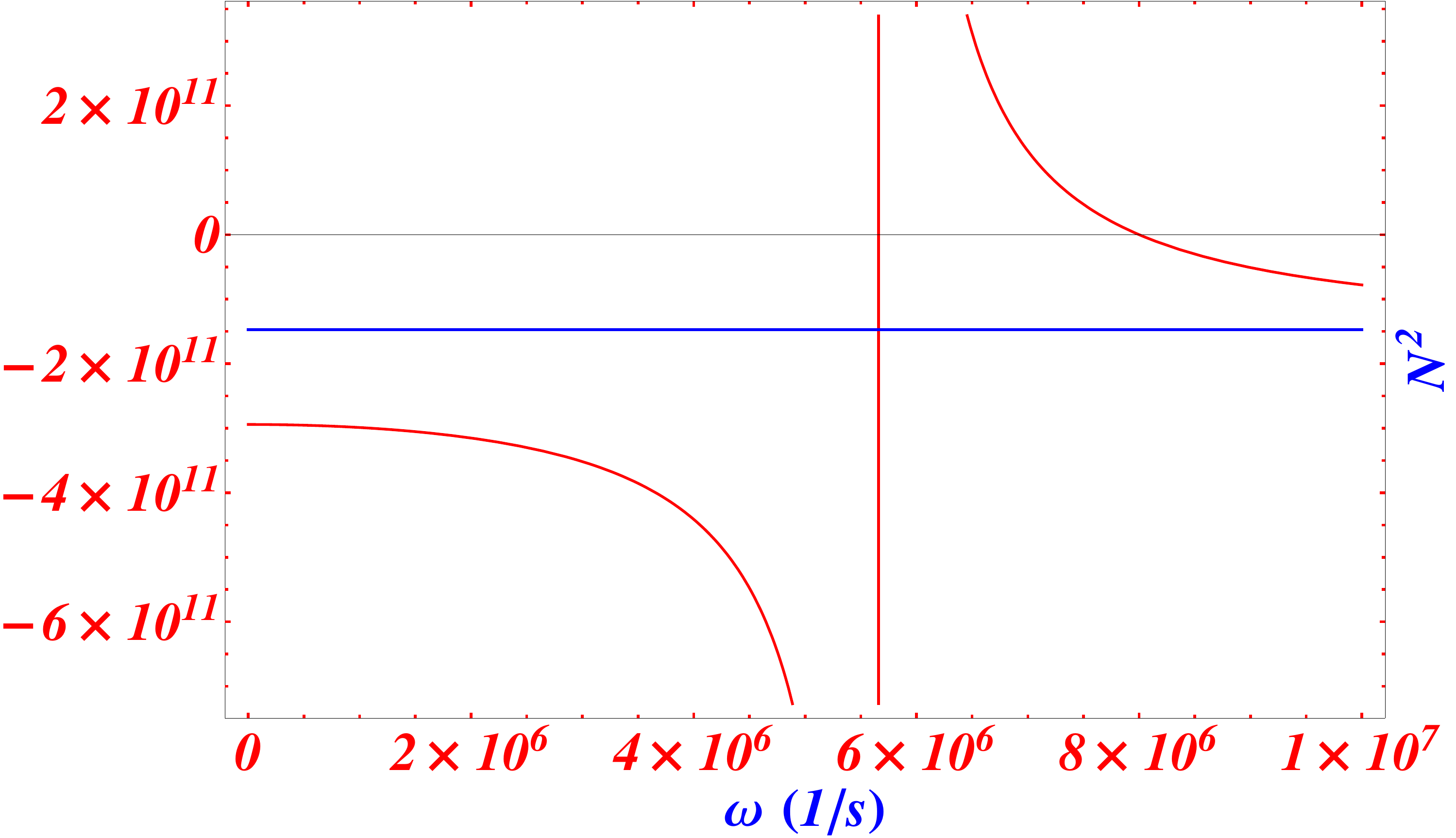} \\\end{tabular}
\caption{{\em The figure shows the corresponding classical refractive index (\ref{EXCH_SEC12}) in the limit of the long wavelength limit $kr_{D}\ll1$, where the Coulomb exchange interactions are dominate $\gamma_{ex}^{2}>\upsilon_{Fe}^{2}$. System parameters are assumed to be
as follows: $n_0\simeq 10^{15} sm^{-3}$, $B_0=5\cdot10^3$ G. The {\color{blue}blue branch} is the index {\color{blue}$c^2/\gamma_{ex}^{2}.$} }
    }\label{N1} \end{figure}

\begin{figure} [h]
   \centering\begin{tabular}{c}
     \includegraphics [scale=0.27]{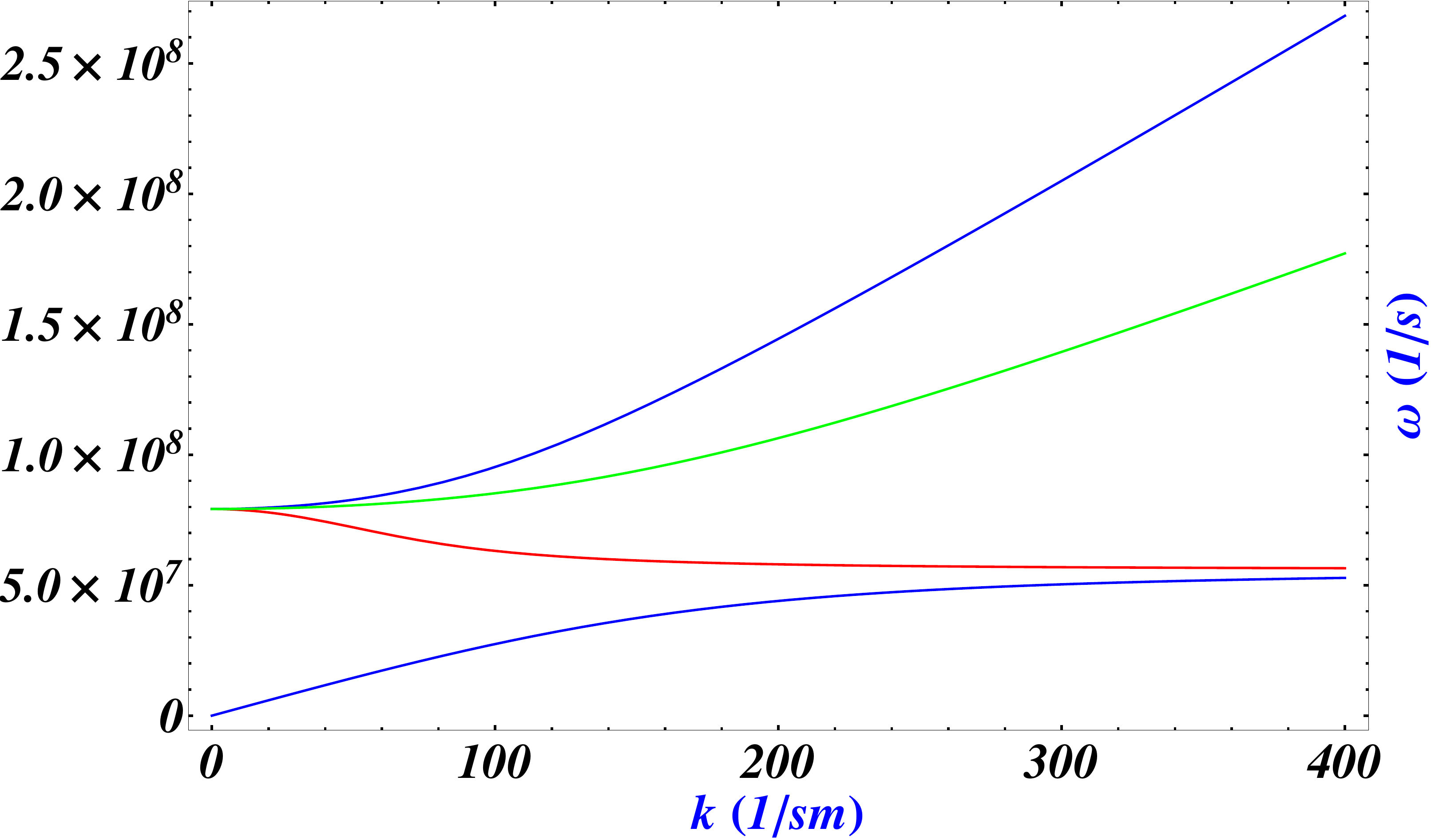} \end{tabular}
\caption{{\em The figure shows the dispersion characteristic of the
slow  and fast  magneto-sonic  waves, which is described by the equation (\ref{EXCH_SEC longit waves in magn field}) in  the long wavelength limit $kr_{D}\ll1$, $n_0\simeq 10^{23} sm^{-3}$, $B_0=5\cdot10^4$ G. The {\color{red}red branch} characterizes the dispersion of  total polarized system $\eta=1$, the {\color{blue}blue mode} $\eta=0.3$ and {\color{green}green mode} $\eta=0.2$ present the dispersion properties of  partially polarized systems.}
    }\label{ex1} \end{figure}

Using the definition (\ref{v}) the Coulomb exchange pressure can be important for the slow and  fast magneto-sonic wave, see Fig. (\ref{ex1}).
%\begin{figure} [h]
 %  \centering\begin{tabular}{c}
  %   \includegraphics [scale=0.32] {ex2.pdf} \\\end{tabular}
%\caption{{\em The figure shows the dispersion characteristic of the
% fast magneto-sonic  waves, which is described by the equation (\ref{EXCH_SEC longit waves in magn field}) in the limit of the long wavelength limit $kr_{D}\ll1$, where the exchange effects are taken into account.}
 %   }\label{ex2} \end{figure}

\section{\label{sec:level6} The low-frequency electromagnetic oscillations in the magnetized plasma}

Here we discuss low-frequencies electromagnetic waves at $T_{e}\gg T_{i}$ under influence of the Coulomb exchange interaction.

Dispersion equation existing in the case under consideration is rather huge. Thus we do not present it here. Nevertheless, we present description of limit cases.
In low frequency limit, it corresponds to the long wavelength limit $k\rightarrow0$, we have
the dispersion of the slow and fast magneto-sonic waves        \begin{widetext}
  \begin{figure} [h]
   \centering\begin{tabular}{c}
     \includegraphics [scale=0.7] {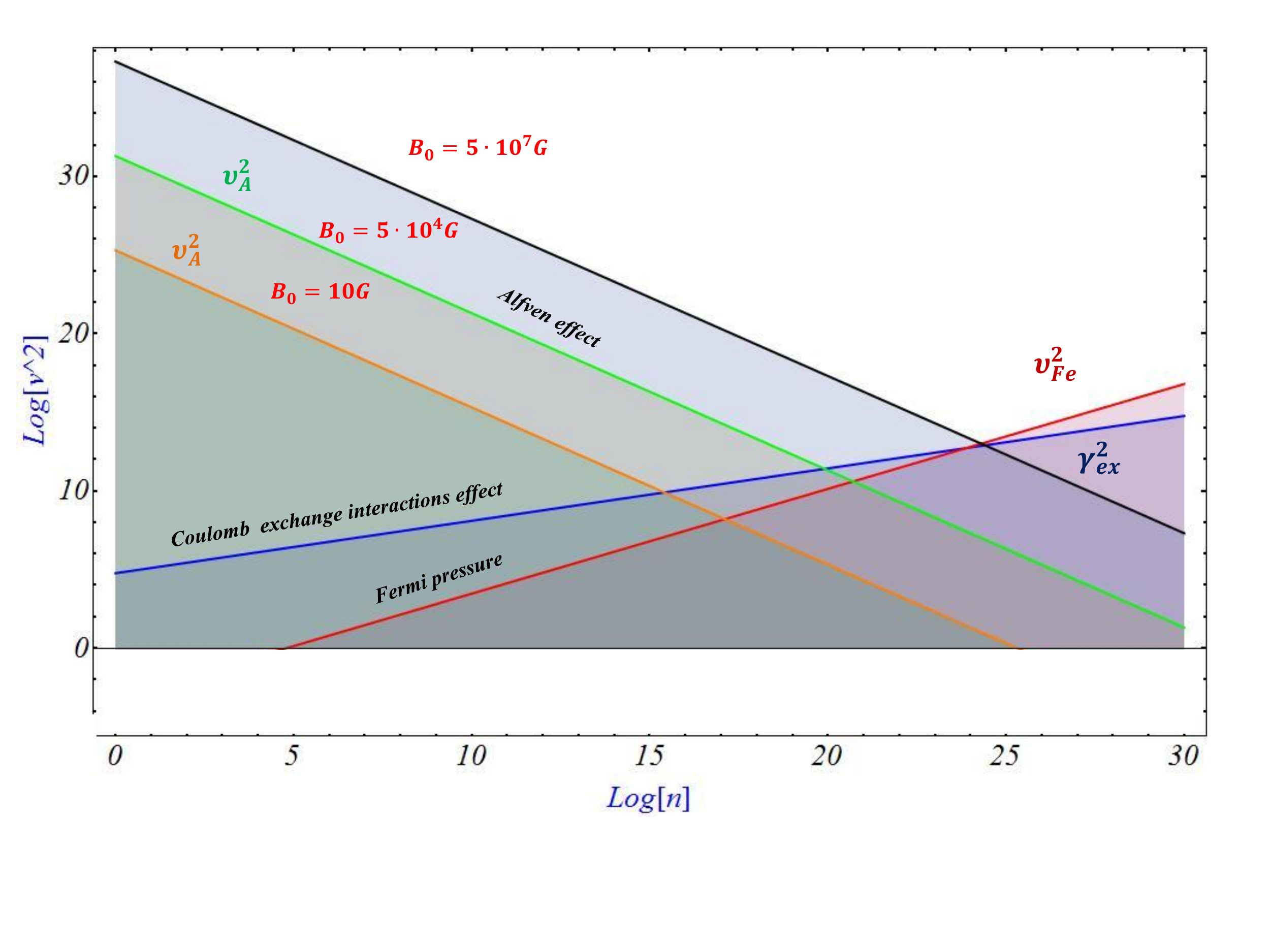} \\\end{tabular}
\caption{{\em A figure illustrating regions of importance in parameter space for various quantum plasma effects of the longitudinal wave propagating perpendicular to the external   magnetic field (\ref{w20}). The orange, green and black branches describe the  Alfv´en
regime.}
    }\label{k20} \end{figure} \end{widetext}
\begin{equation}\label{EXCH_SEC13} \omega=kv_{A}\cos\theta\end{equation}
and
\begin{equation}\label{EXCH_SEC14} \omega=kv_{\pm},\end{equation}
where     the Alfven velocity    is
\begin{equation}\label{EXCH_SEC15} v_{A}=c\frac{\Omega_{i}}{\omega_{Li}}=\frac{B_{0}}{\sqrt{4\pi n_{0}m_{i}}}\end{equation}
and                                                                                                       \begin{equation}\label{EXCH_SEC16} v_{\pm}^{2}=\frac{1}{2}(v_{A}^{2}+v_{s}^{2})\pm\frac{1}{2}\sqrt{(v_{A}^{2}+v_{s}^{2})^{2}-4v_{A}^{2}v_{s}^{2}\cos^{2}\theta}\end{equation}

For the wave propagating parallel to the external magnetic field  $\cos\theta\simeq0$ the dispersion low (\ref{EXCH_SEC14}) has the form
            \begin{equation}\label{DL} \omega_{\pm}=k \begin{cases}
v_{A},  \\
v_{s}.
\end{cases}\end{equation}

Lets consider the longitudinal wave propagating perpendicular to the external   magnetic field. The dispersion low  (\ref{EXCH_SEC14}) takes the form

 \begin{equation}  \label{w20}   \omega^2=k^2\Biggl(\upsilon^2_{A}+\upsilon^2_{i}-\xi_{3D}\frac{e^2}{m_i}\sqrt[3]{\frac{3}{\pi}}\sqrt[3]{n_{0e}}
                           \end{equation}
                                       \[ \qquad\qquad\qquad +\vartheta_{3D}\frac{(3\pi^2)^{2/3}\hbar^2n_{oe}^{2/3}}{3m_em_i}+\frac{\hbar^2k^2}{4m_em_i}\Biggr),\]
                          \begin{equation}  \vartheta_{3D}=\frac{1}{2}[(1+\eta)^{5/3}+(1-\eta)^{5/3}], \end{equation} \[ \xi_{3D}=[(1+\eta)^{4/3}-(1-\eta)^{4/3}].  \]
         The  figure (\ref{k20}) shows regions of importance in parameter space for various quantum plasma effects $\eta=1$.  The Fermi-pressure  $\sim\upsilon^2_{Fe}\sim n_{oe}^{2/3}$
becomes important when the Fermi temperature approaches the thermodynamic temperature, when the plasma concentration $n_0\geq10^{25}$cm$^{-3}$. The
 Alfv´en mode described by the first term $\upsilon^2_{A}$ on the right side of (\ref{w20}). The effects due to the magnetic pressure  depend on the  magnetic field strength $\upsilon^2_{A}\sim n_0^{-1}$. The magnetic pressure effects can be important in regimes of   $n_0\leq10^{20}$cm$^{-3}$ and external magnetic field $B_0\simeq5\cdot10^{4}$G.  The quantum regime correspond to lower temperatures. The Coulomb exchange interactions proportional to $\sim\gamma^2_{ex}\sim n_{0e}$ can be important in regimes of   $10^{20}\leq n_0\leq10^{24}$cm$^{-3}$.  The Coulomb exchange  force does not provide a
stabilizing mechanism.   The exchange pressure $\sim\gamma^2_{ex}=\xi_{3D}\sqrt[3]{\frac{3}{\pi}}\sqrt[3]{n_{0e}}e^2/m_i$ is the negative pressure term and therefore the source of the instability.  But for the high magnitude magnetic field $B_0\sim10^7 G$, the instability is stabilized.

\begin{figure} [h]
   \centering\begin{tabular}{c}
     \includegraphics [scale=0.21] {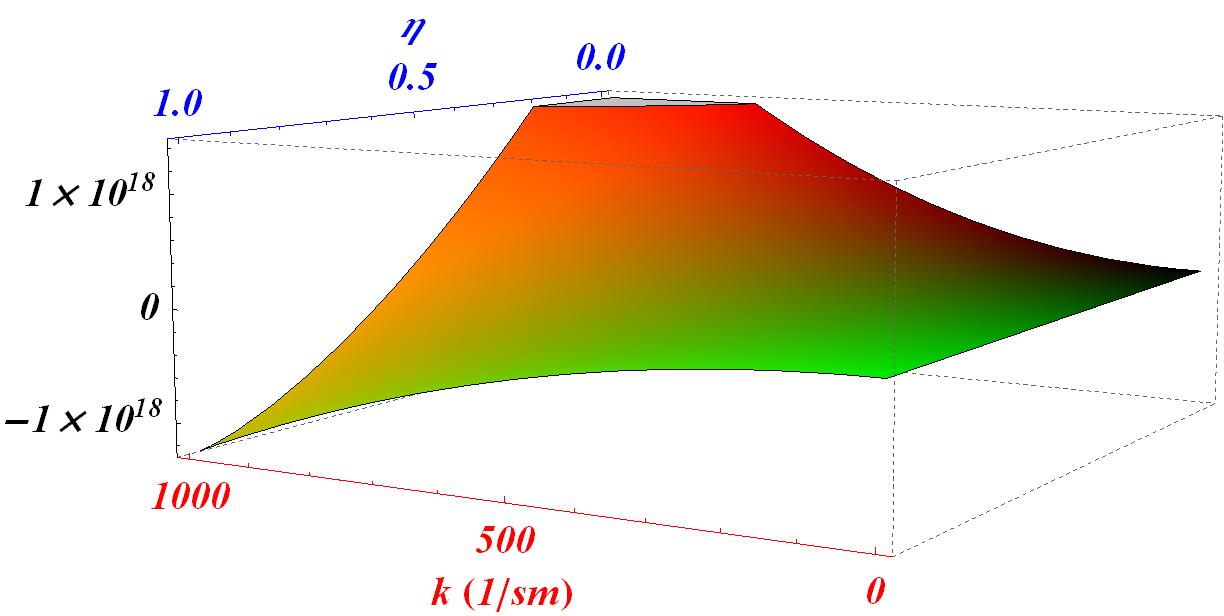} \\\end{tabular}
\caption{{\em The 3D figure shows the dispersion characteristic of the
  magneto-sonic  waves, which is described by the equation (\ref{w20}), where the exchange effects are taken into account. The 3D figure represents the wave frequency $\omega^2(k,\eta)$  as a function of  the wave
vector $k$ and the ratio of
polarizability $\eta$.  The Coulomb exchange interactions and quantum Bohm potential are taken into account and  $n_{0e}\simeq10^{23}cm^{-3}$.}
    }\label{ex3} \end{figure}

Let us consider the small wavelength limit $k\rightarrow\infty$. In this regime
the refractive index and frequency of smallest of three solutions appear as
\begin{equation}\label{EXCH_SEC17} N^{2}=(2N_{A}^{2}+N_{s}^{2}\sin^{2}\theta)\frac{\Omega_{i}^{2}}{\Omega_{i}^{2}\cos^{2}\theta-\omega^{2}},\end{equation}
and
\begin{equation}\label{EXCH_SEC18} \omega(k,\theta)=\Omega_{i}\cos\theta\times\end{equation}
\[\times\Biggl[1-\frac{\Biggl[2\Biggl(-\xi_{3D}\frac{e^2}{m_i}\sqrt[3]{\frac{3}{\pi}}\sqrt[3]{n_{0e}}+\frac{\hbar^2k^2}{4m_em_i}\Biggr)+\sin^{2}\theta v_{A}^{2}\Biggr]\Omega_{i}^{2}}{2k^{2}v_{A}^{2}\Biggl(-\xi_{3D}\frac{e^2}{m_i}\sqrt[3]{\frac{3}{\pi}}\sqrt[3]{n_{0e}} +\frac{\hbar^2k^2}{4m_em_i}\Biggr)}\Biggr],\]
where $N_{A}=c/v_{A}$, $N_{s}=c/v_{s}$. The dispersion of quantum slow magneto-sonic waves in the small wavelength limit $k\rightarrow\infty$ is presented at Fig. (\ref{ex3}).
                \begin{figure} [htbp]
   \centering\begin{tabular}{c}
     \includegraphics [scale=0.25] {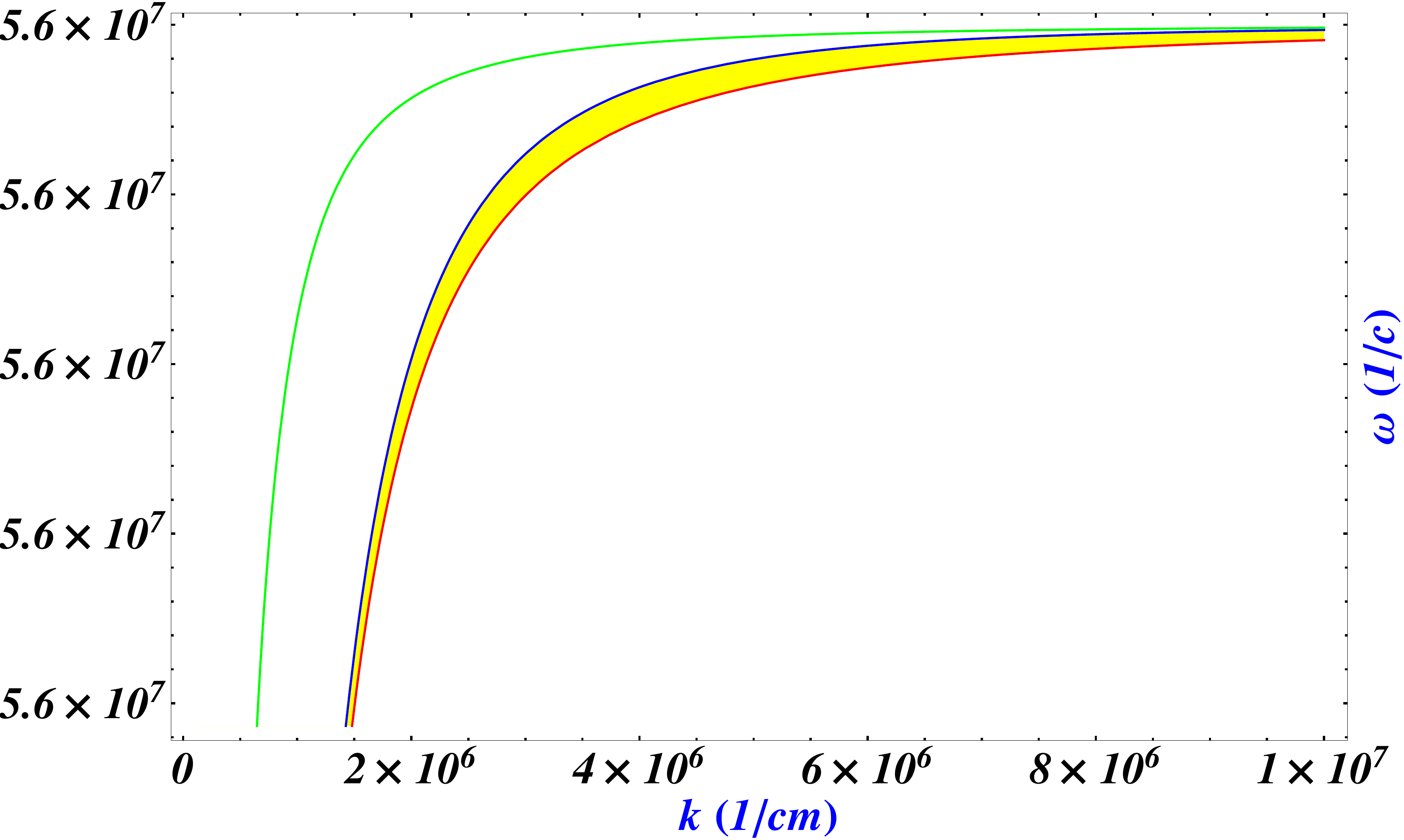} \\\end{tabular}
\caption{{\em The figure shows the dispersion characteristic of the
 slow magneto-sonic  waves, which is described by the equation (\ref{EXCH_SEC18}) in the small wavelength limit $k\rightarrow\infty$, where the exchange effects are taken into account. The {\color{red}red branch} of dispersion represents the classical low, the {\color{blue}blue branch} include the dispersion characteristic of quantum slow magneto-sonic  waves which occurs due to
Bohm quantum potential and the {\color{green}green mode} takes account of Coulomb exchange interactions.}
    }\label{ex3} \end{figure}

\section{Conclusions}

We have briefly described quantum hydrodynamic model for the  magnetized quantum plasmas. In our work
we consider the electron-electron and ion-ion Coulomb exchange interactions. Using QHD equations with Coulomb exchange force field we analyzed elementary excitations
in various physical systems in a linear approximation.  We investigated the  dispersion properties of  the  ion-acoustic waves in non-isothermal magnetized plasmas, the Trivelpiece-Gould waves,  the longitudinal low-frequency ion-acoustic waves, the magneto-sonic waves, high-frequency and low-frequency electron sound tracing contribution of the
exchange interaction. We described the different regimes and showed that the exchanges interactions can lead to instability.

%%%%%%%%%%%%%%%%%%%%%%%%%%%%%%%%%%%%%%%%%%%%%

\begin{acknowledgements}
The authors thank Professor L. S. Kuz'menkov for fruitful discussions.
\end{acknowledgements}

\end{document}